\documentclass[runningheads]{llncs}

\usepackage{array,multirow,graphicx}
\usepackage{graphicx}
\usepackage{epsf} 
\usepackage{psfrag} 
\usepackage{epsfig}
\usepackage{ifthen}
\usepackage{law}
\usepackage{mystyle}
\usepackage{graphics}
\usepackage{times}
\usepackage{cite}
\usepackage{amsbsy}
\usepackage{amsfonts} 
\usepackage{amsmath} 
\usepackage{amssymb} 
\usepackage{multirow}
\usepackage{multicol}
\usepackage{float}
\usepackage{steinmetz}
\usepackage{algorithm} 
\usepackage{algorithmic}
\usepackage{wrapfig}
\usepackage{url}
\usepackage{arydshln}
\usepackage{tabularx}

\input{psfig.sty}
\input{epsf.sty}

\begin{document}

\title{Precise Feature Selection and Case Study of Intrusion Detection in an Industrial Control System (ICS) Environment\thanks{Supported by Cybersecurity Education, Research and Outreach Center (CEROC), as well as Center for Manufacturing Research (CMR), both at Tennessee Tech University.}}
%
%
\author{Terry Guo \and
Animesh Dahal \and
Ambareen Siraj
}
\authorrunning{T. Guo et al.}
\titlerunning{Precise Feature Selection and Case Study ...}
\institute{Tennessee Tech University, Cookeville, TN 38505, USA \\
nguo@tntech.edu; adahal42@students.tntech.edu; ASiraj@tntech.edu
}
\maketitle              

\vspace{-5mm}
\begin{abstract}
This paper presents analytical techniques to improve redundancy and relevance assessment for precise selection of features in practical multi-class raw datasets. We propose a matrix-rank based $k$-medoids algorithm that guarantees to output all independent medoids. The new algorithm uses matrix rank as a robust indicator, while a traditional $k$-medoids algorithm depends on specific datasets and how the distance between any of two features is defined. Another advantage is that the total number of operations in the nested loops is bounded, different from some $k$-medoids algorithms that involve random search. Sparse regression is an efficient tool for feature relevance analysis, but its outcome can depend on what labeled datasets are employed. A compensation method is introduced in this paper to handle the unequality of class-occurrence in a practical raw dataset. To assess the proposed techniques quantitatively, an existing Industrial Control System (ICS) dataset is used to perform intrusion detection. The numerical results generated from this case study validate the effectiveness and necessity of the proposed analytical framework.

\end{abstract}

\section{Introduction}
Accurate selection of the features in an experimental dataset is the key to successful classification. To use the features wisely, it is necessary to identify the ``right'' features  that can lead to reduction in run time and/or improvement of classification performance. The process of selecting a subset of relevant features from a large set of features is called feature selection which can often times yield an efficient learning model \cite{liu2005toward}. As mentioned in \cite{xue2016survey}, feature selection can be used in data from various fields to create a fast and efficient learning model, for example to quickly discover key genes from a large number of candidate genes in biomedical problems \cite{ahmed2013enhanced}, to investigate representative features that describe the dynamic business environment \cite{liu2005toward}, to identify key terms like words or phrases in text mining \cite{aghdam2009text}, and to choose and construct important visual compositions like shape, texture, pixel and color in image analysis \cite{ghosh2013self}. Similarly, feature selection can be used to build efficient intrusion detection system by selecting most important features \cite{ambusaidi2016building}.

Features can be categorized into three groups: relevant features, irrelevant features and redundant features, note that a relevant feature can be redundant as well. 
It is desirable to identify and eliminate redundant and irrelevant features in a dataset of interest. In general, these issues are related to ``feature selection'' \cite{narendra1977branch,
dash1997feature,
mitra2002unsupervised,
liu2005toward,peng2005feature,shen2008feature,
covoes2009cluster,covoes2009experimental,jaskowiak2010comparative,nie2010efficient,xiang2012discriminative,
cai2013exact,
song2013fast,zhao2013similarity,chandrashekar2014survey,hou2014joint,
peng2015direct,liu2016consensus,ang2016supervised,
peng2017general,gossmann2017sparse,shang2018non}. 
Feature selection enables development of simpler and faster learning algorithms by saving memory and eliminating irrelevant features. The removal or selection of such relevant yet redundant features may lead to sub-optimal or optimal feature subset, making feature selection a tricky task \cite{xue2016survey}. 
There are many existing feature selection methods, and they can be categorized into filters, wrappers, embedded and others \cite{chandrashekar2014survey,ang2016supervised}. 
However, filter and wrapper based techniques are the two representative approaches to feature selection \cite{xue2016survey}. The wrapper approach includes a classification/learning algorithm in the feature subset evaluation step which is used to evaluate the goodness of the selected features. Whereas, the filter approach is not dependent on any classification algorithm. Generally, filter approaches tend to be computationally less expensive compared to wrapper approaches \cite{dash1997feature,liu2009manipulating,liu2010feature}. Our technique is a filter based feature selection approach which is suitable for effective and efficient dimensionality reduction in a high dimensional dataset. 
It needs to be pointed out that in literature the two issues related to feature selection, redundancy and relevance, may not be handled at the same time. In \cite{zhao2010efficient,zhao2013similarity} both relevance and redundancy are taken into account in spectral feature selection at relatively high computation. In this paper we consider supervised feature selection and deal with the problem by conducting two separated tasks: redundancy analysis and relevance analysis. 

The fundamental idea for redundancy analysis is distance (or similarity) based clustering. In general, $k$-medoids clustering with predefined distance measure can partition features into clusters based on the distances between them \cite{
reynolds2004application,
park2006k,park2009simple,covoes2009cluster,covoes2009experimental,jaskowiak2010comparative
}. However, the performance of $k$-medoids clustering depends on what specific dataset is used and how a distance measure is defined \cite{jaskowiak2010comparative,jain1998algorithms}. In addition, the number ($k$) of clusters is a critical predetermined parameter to most clustering algorithms, but it is not straightforward to determine its value. Simplified Silhouette Filter (SSF) \cite{mitra2002unsupervised,covoes2009cluster,jaskowiak2010comparative} is a clustering method that does not need to know the number of clusters in prior. However, it is found that this method is computationally expensive and not quite robust. 
In this paper we propose an alternative clustering technique that relies on measuring matrix rank thus is more robust. The proposed feature matrix rank based $k$-medoids clustering algorithm does not need an exhaustive search to determine parameter $k$. Moreover, the algorithm has a bounded complexity.

A feature, even if it is not redundant, could be irrelevant to a classification task. Evaluating feature relevance is as important as assessing feature redundancy in feature selection. Recently, sparse regression based feature relevance analysis has drawn attention \cite{nie2010efficient,xiang2012discriminative,
cai2013exact,
peng2015direct,peng2017general,gossmann2017sparse}. Algorithms in this subset belong to embedded feature selection category and typically exhibit both efficiency and tractability. For a given dataset with labels, one hidden parameter is the class occurrence, i.e., the number of instants that are associated with a particular class. As verified by experiment, class occurrences do affect analysis result. We introduce a compensation method that can be integrated with existing sparse regression framework for relevance analysis.


Industrial Control Systems (ICSs) of the past have been shielded from network intrusions by means of an ``air gap'' separating the system from the open internet. However, this protection is no longer universally present in modern networked ICSs. There has been a growing demand for designing protection mechanisms against various attacks on the ICSs, and intrusion detection is one of such mechanisms. The proposed feature selection techniques are examined by using a case study of ICS intrusion detection.

\textbf{Major Contributions} in this work include:
\begin{enumerate}
	\item Proposal of a matrix-rank-preserving $k$-medoids algorithm which is more robust and has a bounded complexity;
	\item Proposal of a class-occurrence compensation technique integrated with the $l_{2,1}$-norm minimization framework to ensure fairness of feature relevance analysis.
	\item Experimental validation of the proposed techniques.
\end{enumerate}

The rest of the paper is organized as follows. The feature redundancy analysis including a matrix-rank-preserving $k$-medoids algorithm is provided in the next section. Section III introduces the compensation for fair assessment with the sparse regression based feature relevance analysis. 
A case study of ICS intrusion detection is given in Section IV to generate numerical results and validate the proposed techniques. Section V summarizes our work and presents some remarks.

\begin{figure}
	\centering
		\includegraphics[width=0.8\textwidth]{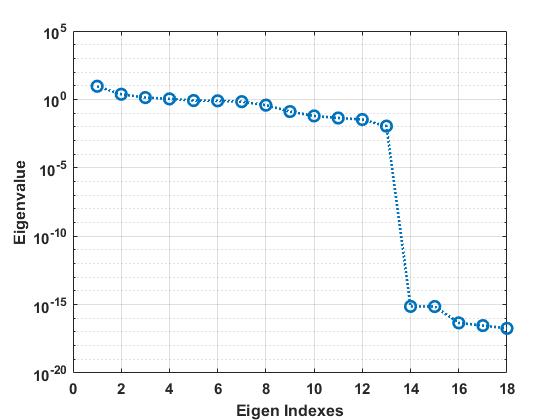}
\caption{Eigen spectrum of the water storage tank dataset \cite{morris11industrial}.}
\label{eigen_spectrum}
\end{figure}

\vspace{-8mm}
\section{Matrix-Rank Based Redundant Feature Identification}

In this paper we propose an alternative clustering technique that relies on measuring matrix rank thus is more robust and accurate. 
In the analysis below a given dataset is represented as either an $m\times n$ matrix $\boldsymbol{F}=\left(\boldsymbol{f}_1, \boldsymbol{f}_2, \cdots, \boldsymbol{f}_m\right)^T\in\mathbb{R}^{m\times n}$ or an $m$-member set ${\cal F}=\left\{\boldsymbol{f}_1, \boldsymbol{f}_2, \cdots, \boldsymbol{f}_m\right\}$, where each member represents a feature, $m$ is the number of features and $n$ is the number of instants. The rank of matrix $\boldsymbol{F}\boldsymbol{F}^T/n$ (the sample covariance matrix of the feature dataset) tells how many significant eigen modes $\boldsymbol{F}$ contains. For instance, from the eigenvalue spectrum (shown in Fig.\ref{eigen_spectrum}) of the water tank data matrix we can say that all the information embedded in the feature matrix can possibly be represented by as less as 13 independent features. 

\vspace{-5mm}
\begin{algorithm}[H] 
\caption{Matrix-rank-preserving $k$-medoids algorithm} 
\label{algorithm}  
{\bf Inputs:} data matrix $\boldsymbol{F}$. \\
{\bf Initialization:} ${\cal C}_1={\cal C}_2=\cdots={\cal C}_m=\Phi$; $\boldsymbol{F_0}=\boldsymbol{0}$; $\boldsymbol{S}=\boldsymbol{F}$; $k=0$.\\
{\bf Result:} $k$ clusters and $k$ medoids. \\
{\bf Phase-1:} Find all $k$ clusters ${\cal C}_j,\;j=1,2,3,\cdots,k$. \\
\vspace{-4mm}
\begin{algorithmic}
\WHILE {$\boldsymbol{S}$ is not empty}
\item $k \Leftarrow k+1$; \\
\item remove one row from $\boldsymbol{S}$ and denote it by $\boldsymbol{s}_0$; 
\item add $\boldsymbol{s}_0$ to cluster ${\cal C}_k$; \\
\item $\boldsymbol{F_0} \Leftarrow \boldsymbol{F_0} \boxplus \boldsymbol{s}_0$; \\
\item $i \Leftarrow 1$; \\
\item $len \Leftarrow$ number of row in $\boldsymbol{S}$; \\
\FOR {$r=1$ to $len$}
\item take one row from $\boldsymbol{S}$ and denote it by $\boldsymbol{s}_i$;
\IF {$rank(\boldsymbol{F_0})=rank(\boldsymbol{F_0} \boxplus \boldsymbol{s}_i)$}
\item add $\boldsymbol{s}_i$ to cluster ${\cal C}_k$; \\
\item remove one row from $\boldsymbol{S}$; \\
\ELSE
\item $i \Leftarrow i+1$; \\
\ENDIF
\ENDFOR
\ENDWHILE
\end{algorithmic}
{\bf Phase-2:} Determine $k$ medoids.
\begin{algorithmic}
\FOR {$j=1$ to $k$}
\IF {$|{\cal C}_j|\ge 3$}
\item choose a member from ${\cal C}_j$ as the cluster medoid such that the sum of its distances to its neighbors is minimal; \\
\ELSE
\IF {$|{\cal C}_j|=2$}
\item randomly choose one of the two members in ${\cal C}_j$ as the cluster medoid; \\
\ELSE
\item the sole member of ${\cal C}_j$ is the cluster medoid; \\
\ENDIF
\ENDIF
\ENDFOR
\item * Symbol ``$\boxplus$'' represents attaching a row to a matrix.
\end{algorithmic}
\end{algorithm}
\vspace{-5mm}

The proposed algorithm is shown in {\bf Algorithm \ref{algorithm}} and it relies on the following facts. Let $\tilde{\boldsymbol{F}}$ be a $p\times n$ matrix that contains $p(< m, n)$ rows, and $\tilde{\boldsymbol{F}}_{(i)}$ be a $(p+1)\times n$ matrix that contains all rows of $\tilde{\boldsymbol{F}}$ and an additional row $\boldsymbol{f}_i$. Condition $rank(\tilde{\boldsymbol{F}}) = rank(\tilde{\boldsymbol{F}}_{(i)})$ is satisfied, if and only if $\boldsymbol{f}_i$ depends on any of rows in $\tilde{\boldsymbol{F}}$. The algorithm does not require the parameter $k$ to be set in advance. Another advantage of this algorithm is that the total number of operations in the nested loops is bounded, while many $k$-medoids algorithms do not have bounded complexities because of random search. The bound of loop operations in Phase-1 is $(m-1)+(m-1)+\cdots+1=m(m-1)/2 \sim O(m^2)$.


The medoid selection method (Phase-2) used in the algorithm is based on a distance metric defined as the total distance from a reference feature to all its neighbors, though there can be other criteria for medoid selection. Other than the medoids that have been recognized, all the rest of features are redundant.

\section{Feature Relevance Analysis For Practical Datasets}
Among many feature relevance analysis techniques are those based on sparse regression 
which are attractive in terms of computation and traceability \cite{nie2010efficient,xiang2012discriminative,
cai2013exact,
peng2015direct,peng2017general,gossmann2017sparse}. In particular, the techniques using joint $l_{2,1}$-norms minimization \cite{nie2010efficient} are especially interesting to us for its simplicity and efficiency. 

\subsection{Measuring Feature Relevance Based On $l_{2,1}$-norm Minimization}
The goal is to find a weighting matrix $\boldsymbol{W}$ in a supervised learning manner. We adopt the framework used in \cite{nie2010efficient} and the problem is formulated as follows. 

Let $c$ be the number of classes. Define the weighting matrix $\boldsymbol{W}=\left(\boldsymbol{w}_1, \boldsymbol{w}_2, \cdots, \boldsymbol{w}_m\right)^T \in \mathbb{R}^{m\times c}$ and its extended version $\hat{\boldsymbol{W}}
\in \mathbb{R}^{d\times c}$, 
\begin{eqnarray}
	\hat{\boldsymbol{W}} &=&
\left(\begin{array}{c}
	\boldsymbol{W} \\ \hdashline[2pt/2pt]
	\hat w_{d,1}, \cdots, \hat w_{d,c}
	\end{array}\right) ,
\end{eqnarray}
with 
$d=m+1$. The value of $\hat{\boldsymbol{W}}$ will be determined later. Extend the data matrix $\boldsymbol{F}$  into ${\boldsymbol{X}}=\left({\boldsymbol{x}}_1, {\boldsymbol{x}}_2, \cdots, {\boldsymbol{x}}_n\right) \in\mathbb{R}^{d\times n} $ by adding an all-one row at the bottom of $\boldsymbol{F}$,
\begin{eqnarray}
	\boldsymbol{X} &=& 
\left(\begin{array}{c}
	\boldsymbol{F} \\ \hdashline[2pt/2pt]
	1, \cdots, 1
	\end{array}\right) ,
\end{eqnarray}
Assume the dataset comes with $n$ label samples denoted by $a_1, a_2, \cdots, a_n \in \{1, 2, \cdots, c\}$. Denote the class label matrix by $\boldsymbol{Y}=\left(\boldsymbol{y}_{a_1}, \boldsymbol{y}_{a_2}, \cdots, \boldsymbol{y}_{a_n}\right) \in\mathbb{R}^{c\times n}$, where a column vector $\boldsymbol{y}_j=(0, \cdots,0,  1, 0, \cdots, 0)^T$ contains $c-1$ zero-valued entries and a sole one-valued entry at the $j$-th position associated with the class $j$. To find $\hat{\boldsymbol{W}}$, the regression (minimization) problem is
\begin{eqnarray}
\label{minimization}
	\min_{\hat{\boldsymbol{W}}} \sum_{i=1}^n \left\rvert\left\rvert \hat{\boldsymbol{W}}^T \boldsymbol{x}_i - \boldsymbol{y}_i \right\rvert\right\rvert_2 + \gamma \sum_{j=1}^d \left\rvert\left\rvert \hat{\boldsymbol{w}}_j \right\rvert\right\rvert_2
\end{eqnarray}
where $\hat{\boldsymbol{w}}_j$ is the $j$-th row of $\hat{\boldsymbol{W}}$, $\sum_{i=j}^d \left\rvert\left\rvert \hat{\boldsymbol{w}}_j \right\rvert\right\rvert_2$ is the regularization term, 
and $\gamma$ is a constant for tuning the regularization's influence. The problem \eqref{minimization} can be efficiently solved using the algorithm described in \cite{nie2010efficient} (refer to the reference for the analysis and proof). The first $m$ rows of $\hat{\boldsymbol{W}}$, i.e., $\boldsymbol{W}$, is the outcome we expect. Each of $m\cdot c$ entries of $\boldsymbol{W}$ reflects how relevant a feature is to a class.

With $\boldsymbol{W}$ we can also evaluate how important an individual feature is to the overall classification. By adopting the way used in \cite{xiang2012discriminative}, the total relevance of the $j$-th feature can be calculated by
\begin{eqnarray}
\label{relevance}
	\bar w_j &= \left\rvert\left\rvert \boldsymbol{w}_j  \right\rvert\right\rvert_2, j=1,2,\cdots,m
\end{eqnarray}

\subsection{Class-Occurrence Compensation}
The relevance analysis method presented in the last subsection will not work well if no proper compensation for class occurrence is made. Let $n_l$ be the number of instants associated with class $l,\;l=1,2,3,\cdots,c$. Consider an ideal case that $n_l=n/c,\;l=1,2,3,\cdots,c$, i.e., equal occurrence for all $c$ classes, we first apply Z-score normalization to the feature dataset and then calculate the weighting matrix $\boldsymbol{W}$. In this process all classes are represented equally, which is necessary for a fair analysis. However, equal occurrence does not hold in general, thus certain compensations are needed in order to obtain an unbiased analysis result.

In dataset normalization phase, we need to determine the mean $\mu_j$ and standard deviation $\sigma_j$ for each feature in Z-score normalization: $f_{j,i} \Leftarrow (f_{j,i}-\mu_j)/\sigma_j, \;j=1,2,\cdots,m, \;i=1,2,\cdots,n$. $\mu_j$ and $\sigma_j$ are given by
\begin{eqnarray}
\label{z-score}
	\mu_j &=& \frac{1}{c} \sum_{l=1}^c \frac{1}{n_l}\sum_{i=1}^{n} \boldsymbol{1}_l(f_{j,i}) f_{j,i} , \\
	\sigma_j &=& \frac{1}{c} \sum_{l=1}^c \frac{1}{n_l}\sum_{i=1}^{n} \boldsymbol{1}_l(f_{j,i}) (f_{j,i} - \mu_j)^2 , \\
	&& j=1,2,\cdots,m \nonumber
\end{eqnarray}
where $\boldsymbol{1}_l(f_{j,i})$ is an indicator function defined as
\begin{eqnarray}
\label{indicator_function}
	\boldsymbol{1}_l(f_{j,i}) &=& \left\{ \begin{array}{ll}
		1, & $if $ f_{j,i} $ belongs to  class $ l, \\
		0, & $if $ f_{j,i} $ does not belongs to  class $ l, \end{array}\right. \\
	&&\hspace{-15mm} j=1,2,\cdots,m, \; l=1,2,\cdots,c, \; i=1,2,\cdots,n \nonumber
\end{eqnarray}

Certain compensation needs to be made in the phase of $l_{2,1}$-norm minimization as well, and \eqref{minimization} can be extended into the following format:
\begin{eqnarray}
\label{minimization1}
	\min_{\hat{\boldsymbol{W}}} \frac{n}{c} \sum_{l=1}^c \frac{1}{n_l}\sum_{i=1}^n \boldsymbol{1}_l({\boldsymbol{x}}_i) \left\rvert\left\rvert \hat{\boldsymbol{W}}^T \boldsymbol{x}_i - \boldsymbol{y}_i \right\rvert\right\rvert_2 + \gamma \sum_{j=1}^d \left\rvert\left\rvert \hat{\boldsymbol{w}}_j \right\rvert\right\rvert_2
\end{eqnarray}
To use the algorithm developed in \cite{nie2010efficient}, we can convert $\boldsymbol{x}_i$ and $\boldsymbol{y}_i$ into $\tilde{\boldsymbol{x}}_i$ and $\tilde{\boldsymbol{y}}_i$, respectively, using the following formulas:
\begin{eqnarray}
\label{conversion}
	\tilde{\boldsymbol{x}}_i &=& \frac{n}{c n_l}\boldsymbol{x}_i, \text{ if } \boldsymbol{1}_l({\boldsymbol{x}}_i)=1, \nonumber\\
	\tilde{\boldsymbol{y}}_i &=& \frac{n}{c n_l}\boldsymbol{y}_i, \text{ if } \boldsymbol{1}_l({\boldsymbol{y}}_i)=1, \\
	&& \hspace{-15mm} l=1,2,\cdots,c, \; i=1,2,\cdots,n \nonumber
\end{eqnarray}
By combining \eqref{minimization1} and \eqref{conversion}, we reach the following optimization which has the same format as \eqref{minimization}:
\begin{eqnarray}
\label{minimization2}
	\min_{\hat{\boldsymbol{W}}} \sum_{i=1}^n \left\rvert\left\rvert \hat{\boldsymbol{W}}^T \tilde{\boldsymbol{x}}_i - \tilde{\boldsymbol{y}}_i \right\rvert\right\rvert_2 + \gamma \sum_{j=1}^d \left\rvert\left\rvert \hat{\boldsymbol{w}}_j \right\rvert\right\rvert_2
\end{eqnarray}

\section{Case Study Of ICS Intrusion Detection}
Precise feature selection can benefit design and evaluation of an Intrusion Detection System (IDS). In this section we use ICS intrusion detection as an example to examine the proposed techniques. Specifically, the water storage tank dataset provided by Morris's group \cite{morris11industrial} is employed to generate numerical results. The dataset includes class 0 for normal situation and classes 1 to 7 representing seven different types of attacks. Intrusion detection is actually multi-class classification and we use partial decision tree based PART classifier in Weka \cite{holmes1994weka,weka_link} to perform the job. After removal of a few constant (zero-variance) features, the remaining 18 features are used for analysis. As shown in Table \ref{table1}, these 18 features belong to three categories, and six of payload features are directly related to physical parameters.

\begin{table*}[!tp]
\caption{18 effective features in two categories.}
\centering
\begin{tabular}{|l|l|c|c|c|}
\hline
\hspace{10mm}Feature & \hspace{25mm}Description & Network & Payload & Physical\\
\hline
 1. command address & Device ID in command packet & \checkmark & & \\
\hline
 2. response address & Device ID in response packet & \checkmark & & \\
\hline
 \multirow{2}{*}{3. response memory} & Memory start position in response & \multirow{2}{*}{\checkmark} & & \\
	& packet & & & \\
\hline
 4. command memory & Number of memory bytes for & \multirow{2}{*}{\checkmark} & & \\
	\hspace{2mm} count & R/W command & & & \\
\hline
 5. response memory & Number of memory bytes for & \multirow{2}{*}{\checkmark} & & \\
	\hspace{2mm} count & R/W response & & & \\
\hline
 6. comm write fun & Value of command function code & & \checkmark & \\
\hline
 7. response write fun & Value of response function code & &\checkmark  & \\
\hline
 \multirow{2}{*}{8. sub function} & Value of sub-function code in & & \multirow{2}{*}{\checkmark}  & \\
	& the command/response & & & \\
\hline
 9. response length & Total length of response packet & \checkmark & & \\
\hline
 10. HH & Value of HH setpoint & & & \checkmark\\
\hline
 11. H & Value of H setpoint & & & \checkmark \\
\hline
 12. L & Value of L setpoint & & & \checkmark \\
\hline
 13. LL & Value of LL setpoint & & & \checkmark \\
\hline
 14. control mode & Automatic, manual or shutdown & &\checkmark  & \\
\hline
 15. pump state & Compressor/pump state & &\checkmark  & \checkmark \\
\hline
 16. crc rate & CRC error rate & \checkmark & & \\
\hline
 17. measurement & Water level & &\checkmark  & \checkmark \\
\hline
 18. time & Time interval between two packets & \checkmark & & \\
\hline
\end{tabular}
\label{table1}
\end{table*}

\begin{figure}
	\centering
		\includegraphics[width=1.2\textwidth]{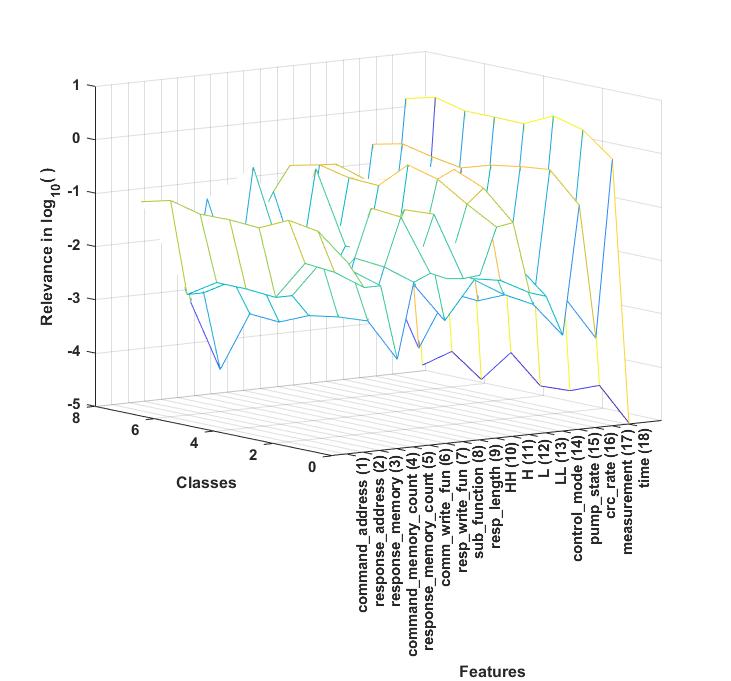}
\caption{Individual relevance.}
\label{individual_relevance}
\end{figure}

\begin{figure}
	\centering
		\includegraphics[width=0.8\textwidth]{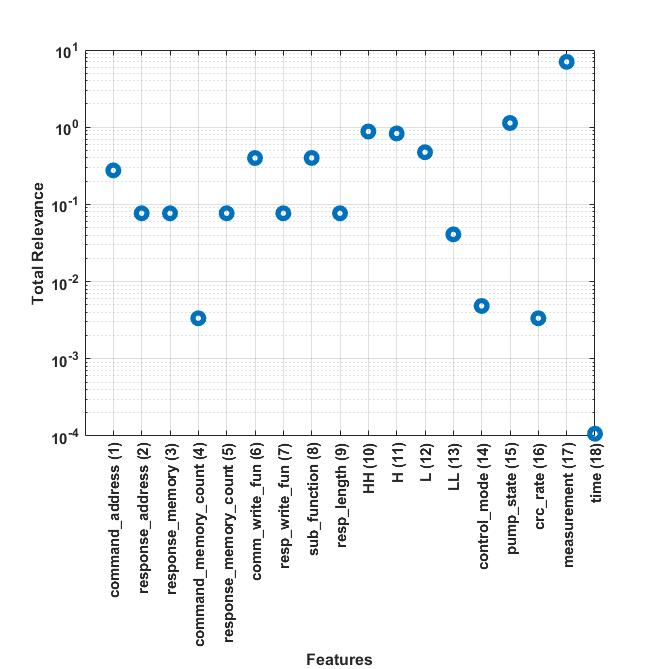}
\vspace{-3mm}
\caption{Total relevance of each feature.}
\label{total_relevance}
\end{figure}

After performing the proposed $k$-medoids algorithm, $k=13$ medoids (primary features) are found. As mentioned above, the matrix rank analysis indicates that this dataset contains 13 effective eigen modes (refer to Fig.\ref{eigen_spectrum}), which implies that, for this particular dataset, each  primary feature corresponds to an effective eigen mode, and all the 13 primary features are independent of each other. The algorithm generates 11 singleton clusters \{1\}, \{6\}, \{8\}, \{10\}, \{11\}, \{12\}, \{13\}, \{14\}, \{15\}, \{17\} and \{18\} along with two non-singleton clusters \{16, 4\} and \{2, 3, 5, 7, 9\} with medoids 4 and 3, respectively. In Fig.\ref{individual_relevance} each data points represents a relevance level of an individual feature with respect to a class, and Fig.\ref{total_relevance} shows overall impact of each feature on all of the classes, where class-occurrence compensation has been performed prior to relevance calculation.

\vspace{-9mm}
\begin{figure}
	\centering
		\includegraphics[width=0.8\textwidth]{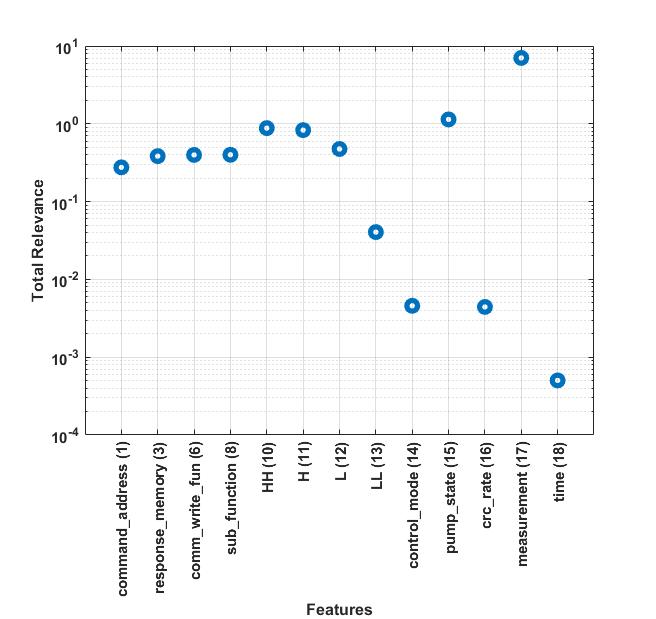}
\vspace{-3mm}
\caption{Total relevance of each feature (only consider 13 independent features).}
\label{total_relevance_13features}
\end{figure}
\vspace{-5mm}

It can be seen in Fig.\ref{total_relevance} that the features belonging to the same cluster exhibit the same relevance level. In practice it is reasonable not to use redundant features, so we should only measure the relevance levels of the 13 independent features that are fed to the classifier. The total relevance of these selected features is shown in Fig.\ref{total_relevance_13features}. 

Table. \ref{table2} shows classification performance for using different feature sets. As expected, it is found that removal of redundant features does not degrade classification performance, and it is even beneficial to eliminate some bad (low-relevance-score) features (e.g., features 4, 13, 14, 16 and 18). It can also be verified that removal of independent and important (high-relevance-score) features can degrade the performance. The 8-feature result shown in the Table \ref{table2} suggests that feature 17 is critical to the classification of the first 3 classes. These observations validate the correctness of the redundancy and relevance analysis.

\begin{table*}[!tp]
\caption{Classification results for (a) all 18 features; (b) 13 features--eliminating 5 redundant features (2, 5, 7, 9, 16); (c) 9 features--eliminating redundant \& bad features (2, 4, 5, 7, 9, 13, 14, 16, 18); (d) 8 features--eliminating feature 17 and 9 redundant \& bad features.}
\centering
\begin{tabular}{|c|c|c|c|c|c|c|c|c|c|c|}
\hline
\multirow{2}{*}{Class} & \multicolumn{2}{|c}{18 features} & \multicolumn{2}{|c|}{13 features} & \multicolumn{2}{c|}{9 features} & \multicolumn{2}{c|}{8 features} \\
\cline{2-9}
& TP & FP & TP & FP &TP & FP &TP & FP \\
\hline
0 & 0.988 & 0.014 & 0.988 & 0.014 & \textbf{0.990} & 0.014 & 1.000 & \textbf{0.346} \\
\hline
1 & 0.977 & 0.000 & 0.977 & 0.000 & \textbf{0.978} & 0.000 & \textbf{0.000} & 0.000 \\
\hline
2 & 0.946 & 0.009 & 0.946 & 0.009 & 0.946 & \textbf{0.007} & \textbf{0.000} & 0.000 \\
\hline
3 & 0.971 & 0.000 & 0.971 & 0.000 & 0.971 & 0.000 & 0.971 & 0.000 \\
\hline
4 & 0.990 & 0.000 & 0.990 & 0.000 & 0.990 & 0.000 & 0.990 & 0.000 \\
\hline
5 & 1.000 & 0.000 & 1.000 & 0.000 & 1.000 & 0.000 & 1.000 & 0.000 \\
\hline
6 & 1.000 & 0.000 & 1.000 & 0.000 & 1.000 & 0.000 & 1.000 & 0.000 \\
\hline
7 & 1.000 & 0.000 & 1.000 & 0.000 & 1.000 & 0.000 & 1.000 & 0.000 \\
\hline
Weighted  & \multirow{2}{*}{0.987} & \multirow{2}{*}{0.010} & \multirow{2}{*}{0.987} & \multirow{2}{*}{0.010} & \multirow{2}{*}{\textbf{0.989}} & \multirow{2}{*}{0.010} & \multirow{2}{*}{\textbf{0.902}} & \multirow{2}{*}{\textbf{0.248}} \\
Average &&&&&&&&\\
\hline
\end{tabular}
\label{table2}
\end{table*}

\begin{figure}
	\centering
		\includegraphics[width=0.8\textwidth]{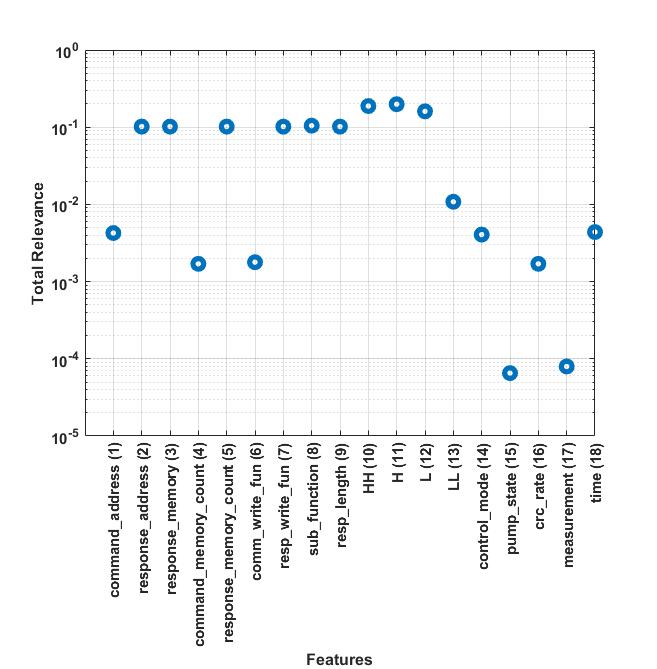}
\caption{Total relevance based on raw dataset without pre-compensation.}
\label{total_relevance1}
\end{figure}

The necessity of class-occurrence compensation can be confirmed experimentally as well. Different from what is shown in Fig.\ref{total_relevance},  a relevance distribution obtained based on the raw dataset without pre-compensation is shown in Fig.\ref{total_relevance1}. It can be verified that removal of the ``bad'' features (e.g., features 15 and 17 are, in fact, very important) suggested by this incomplete analysis can be harmful to the classification task.

It has been seen that, without sacrificing detection accuracy, the intrusion detection complexity can be reduced by using only 9 independent and relevant features. In general, we can have a simpler classifier that uses fewer features with some performance penalties. However, the performance penalties on different classes are not equal. For example, classification result in Table \ref{table3} is obtained by using only 6 features, and the corresponding performance for detecting attacks 1, 2, 3, 4 and 7 is as good as that when more independent and relevant features are utilized. If the attacks corresponding to classes 5 and 6 were not of our interest, we could have designed a light-weight IDS that would have relied only on the 6 features.


\begin{table}[!tp]
\caption{Classification results when using 6 features (1, 3, 10, 11 15, 17).}
\centering
\begin{tabular}{|c|c|c|}
\hline
Class & TP & FP \\
\hline
0 & 0.992 & 0.039 \\
\hline
1 & 0.978 & 0.000 \\
\hline
2 & 0.946 & 0.006 \\
\hline
3 & 0.967 & 0.000 \\
\hline
4 & 0.990 & 0.000 \\
\hline
5 & \textbf{0.000} & 0.000 \\
\hline
6 & \textbf{0.719} & 0.000 \\
\hline
7 & 1.000 & 0.000 \\
\hline
Weighted  & \multirow{2}{*}{0.983} & \multirow{2}{*}{0.028} \\
Average &&\\
\hline
\end{tabular}
\label{table3}
\end{table}


\section{Conclusions}
In this work we have proposed a set of analytical techniques for selecting features efficiently. The matrix rank of feature data is used as a robust indicator for feature clustering. To assess the feature relevance fairly, the unequality of class-occurrence in a practical raw dataset is compensated prior to applying relevance analysis. The compensation idea can be applied to different regression based methods. The effectiveness and necessity of the proposed methods are examined using an existing ICS dataset. One interesting observation from examining the water tank dataset is that some physical features (e.g., features 10, 11,12, 15, 17) can be more important than other types of features. This might be because they are directly related to the physical entities (say, the water level) of interest, suggesting that we could add more sensors to monitor an ICS in order to further improve intrusion detection. Our proposed framework for precise feature selection can help reduce computation of classifiers and guide the design of efficient classification systems, such as an IDS.

\section*{Acknowledgment}
We express our gratitude towards Cybersecurity Education, Research and Outreach Center (CEROC), as well as Center for Manufacturing Research (CMR), both at Tennessee Tech University, for supporting this research. We would also like to acknowledge Dr. Thomas Morris and his colleagues for providing their datasets. 

\bibliographystyle{splncs}
\bibliography{mybib}

\begin{thebibliography}{10}

\bibitem{liu2005toward}
Liu, H., Yu, L.:
\newblock Toward integrating feature selection algorithms for classification
  and clustering.
\newblock IEEE Transactions on knowledge and data engineering \textbf{17}(4)
  (2005)  491--502

\bibitem{xue2016survey}
Xue, B., Zhang, M., Browne, W.N., Yao, X.:
\newblock A survey on evolutionary computation approaches to feature selection.
\newblock IEEE Transactions on Evolutionary Computation \textbf{20}(4) (2016)
  606--626

\bibitem{ahmed2013enhanced}
Ahmed, S., Zhang, M., Peng, L.:
\newblock Enhanced feature selection for biomarker discovery in lc-ms data
  using gp.
\newblock In: Evolutionary Computation (CEC), 2013 IEEE Congress on, IEEE
  (2013)  584--591

\bibitem{aghdam2009text}
Aghdam, M.H., Ghasem-Aghaee, N., Basiri, M.E.:
\newblock Text feature selection using ant colony optimization.
\newblock Expert systems with applications \textbf{36}(3) (2009)  6843--6853

\bibitem{ghosh2013self}
Ghosh, A., Datta, A., Ghosh, S.:
\newblock Self-adaptive differential evolution for feature selection in
  hyperspectral image data.
\newblock Applied Soft Computing \textbf{13}(4) (2013)  1969--1977

\bibitem{ambusaidi2016building}
Ambusaidi, M.A., He, X., Nanda, P., Tan, Z.:
\newblock Building an intrusion detection system using a filter-based feature
  selection algorithm.
\newblock IEEE transactions on computers \textbf{65}(10) (2016)  2986--2998

\bibitem{narendra1977branch}
Narendra, P.M., Fukunaga, K.:
\newblock A branch and bound algorithm for feature subset selection.
\newblock IEEE Transactions on computers \textbf{9}(C-26) (1977)  917--922

\bibitem{dash1997feature}
Dash, M., Liu, H.:
\newblock Feature selection for classification.
\newblock Intelligent data analysis \textbf{1}(3) (1997)  131--156

\bibitem{mitra2002unsupervised}
Mitra, P., Murthy, C., Pal, S.K.:
\newblock Unsupervised feature selection using feature similarity.
\newblock IEEE transactions on pattern analysis and machine intelligence
  \textbf{24}(3) (2002)  301--312

\bibitem{peng2005feature}
Peng, H., Long, F., Ding, C.:
\newblock Feature selection based on mutual information criteria of
  max-dependency, max-relevance, and min-redundancy.
\newblock IEEE Transactions on pattern analysis and machine intelligence
  \textbf{27}(8) (2005)  1226--1238

\bibitem{shen2008feature}
Shen, K.Q., Ong, C.J., Li, X.P., Wilder-Smith, E.P.:
\newblock Feature selection via sensitivity analysis of svm probabilistic
  outputs.
\newblock Machine Learning \textbf{70}(1) (2008)  1--20

\bibitem{covoes2009cluster}
Cov{\~o}es, T.F., Hruschka, E.R., de~Castro, L.N., Santos, {\'A}.M.:
\newblock A cluster-based feature selection approach.
\newblock In: International Conference on Hybrid Artificial Intelligence
  Systems, Springer (2009)  169--176

\bibitem{covoes2009experimental}
Cov{\~o}es, T.F., Hruschka, E.R.:
\newblock An experimental study on unsupervised clustering-based feature
  selection methods.
\newblock In: Intelligent Systems Design and Applications, 2009. ISDA'09. Ninth
  International Conference on, IEEE (2009)  993--1000

\bibitem{jaskowiak2010comparative}
Jaskowiak, P.A., Campello, R.J., Covoes, T.F., Hruschka, E.R.:
\newblock A comparative study on the use of correlation coefficients for
  redundant feature elimination.
\newblock In: Neural Networks (SBRN), 2010 Eleventh Brazilian Symposium on,
  IEEE (2010)  13--18

\bibitem{nie2010efficient}
Nie, F., Huang, H., Cai, X., Ding, C.H.:
\newblock Efficient and robust feature selection via joint $l_{2,1}$-norms
  minimization.
\newblock In: Advances in neural information processing systems. (2010)
  1813--1821

\bibitem{xiang2012discriminative}
Xiang, S., Nie, F., Meng, G., Pan, C., Zhang, C.:
\newblock Discriminative least squares regression for multiclass classification
  and feature selection.
\newblock IEEE transactions on neural networks and learning systems
  \textbf{23}(11) (2012)  1738--1754

\bibitem{cai2013exact}
Cai, X., Nie, F., Huang, H.:
\newblock Exact top-k feature selection via {$l_{2,0}$}-norm constraint.
\newblock In: IJCAI. Volume~13. (2013)  1240--1246

\bibitem{song2013fast}
Song, Q., Ni, J., Wang, G.:
\newblock A fast clustering-based feature subset selection algorithm for
  high-dimensional data.
\newblock IEEE transactions on knowledge and data engineering \textbf{25}(1)
  (2013)  1--14

\bibitem{zhao2013similarity}
Zhao, Z., Wang, L., Liu, H., Ye, J.:
\newblock On similarity preserving feature selection.
\newblock IEEE Transactions on Knowledge and Data Engineering \textbf{25}(3)
  (2013)  619--632

\bibitem{chandrashekar2014survey}
Chandrashekar, G., Sahin, F.:
\newblock A survey on feature selection methods.
\newblock Computers \& Electrical Engineering \textbf{40}(1) (2014)  16--28

\bibitem{hou2014joint}
Hou, C., Nie, F., Li, X., Yi, D., Wu, Y.:
\newblock Joint embedding learning and sparse regression: A framework for
  unsupervised feature selection.
\newblock IEEE Transactions on Cybernetics \textbf{44}(6) (2014)  793--804

\bibitem{peng2015direct}
Peng, H., Fan, Y.:
\newblock Direct {$l_{2,p}$}-norm learning for feature selection.
\newblock arXiv preprint arXiv:1504.00430 (2015)

\bibitem{liu2016consensus}
Liu, H., Shao, M., Fu, Y.:
\newblock Consensus guided unsupervised feature selection.
\newblock In: AAAI. (2016)  1874--1880

\bibitem{ang2016supervised}
Ang, J.C., Mirzal, A., Haron, H., Hamed, H.N.A.:
\newblock Supervised, unsupervised, and semi-supervised feature selection: a
  review on gene selection.
\newblock IEEE/ACM transactions on computational biology and bioinformatics
  \textbf{13}(5) (2016)  971--989

\bibitem{peng2017general}
Peng, H., Fan, Y.:
\newblock A general framework for sparsity regularized feature selection via
  iteratively reweighted least square minimization.
\newblock In: AAAI. (2017)  2471--2477

\bibitem{gossmann2017sparse}
Gossmann, A., Cao, S., Brzyski, D., Zhao, L.J., Deng, H.W., Wang, Y.P.:
\newblock A sparse regression method for group-wise feature selection with
  false discovery rate control.
\newblock IEEE/ACM Transactions on Computational Biology and Bioinformatics
  (2017)

\bibitem{shang2018non}
Shang, R., Wang, W., Stolkin, R., Jiao, L.:
\newblock Non-negative spectral learning and sparse regression-based dual-graph
  regularized feature selection.
\newblock IEEE transactions on cybernetics \textbf{48}(2) (2018)  793--806

\bibitem{liu2009manipulating}
Liu, H., Zhao, Z.:
\newblock Manipulating data and dimension reduction methods: Feature selection.
\newblock In: Encyclopedia of Complexity and Systems Science.
\newblock Springer (2009)  5348--5359

\bibitem{liu2010feature}
Liu, H., Motoda, H., Setiono, R., Zhao, Z.:
\newblock Feature selection: An ever evolving frontier in data mining.
\newblock In: Feature Selection in Data Mining. (2010)  4--13

\bibitem{zhao2010efficient}
Zhao, Z., Wang, L., Liu, H.,  et~al.:
\newblock Efficient spectral feature selection with minimum redundancy.
\newblock In: AAAI. (2010)  673--678

\bibitem{reynolds2004application}
Reynolds, A.P., Richards, G., Rayward-Smith, V.J.:
\newblock The application of k-medoids and {PAM} to the clustering of rules.
\newblock In: International Conference on Intelligent Data Engineering and
  Automated Learning, Springer (2004)  173--178

\bibitem{park2006k}
Park, H.S., Lee, J.S., Jun, C.H.:
\newblock A k-means-like algorithm for k-medoids clustering and its
  performance.
\newblock Proceedings of ICCIE (2006)  102--117

\bibitem{park2009simple}
Park, H.S., Jun, C.H.:
\newblock A simple and fast algorithm for k-medoids clustering.
\newblock Expert systems with applications \textbf{36}(2) (2009)  3336--3341

\bibitem{jain1998algorithms}
Jain, A.K., Dubes, R.C.:
\newblock Algorithms for clustering data.
\newblock Upper Saddle River, NJ, USA: Prentice-Hall, Inc. (1988)

\bibitem{morris11industrial}
Morris, T., Gao, W.:
\newblock Industrial control system network traffic data sets to facilitate
  intrusion detection system research.
\newblock Critical infrastructure protection VIII—8th IFIP WG \textbf{11}
  (2014)  17--19

\bibitem{holmes1994weka}
Holmes, G., Donkin, A., Witten, I.H.:
\newblock Weka: A machine learning workbench.
\newblock In: Intelligent Information Systems, 1994. Proceedings of the 1994
  Second Australian and New Zealand Conference on, IEEE (1994)  357--361

\bibitem{weka_link}
{GNU G}eneral~{P}ublic {L}icense:
\newblock Weka 3: Data mining software in {J}ava

\end{thebibliography}
%

\end{document}